# Ubiquitous HealthCare in Wireless Body Area Networks - A Survey

N. Javaid[1], N. A. Khan[1], M. Shakir[1], M. A. Khan[1], S. H. Bouk[1], Z. A. Khan[2]

[1]*COMSATS Institute of Information Technology, Islamabad, Pakistan.*

*[nadeemjavaid@comsats.edu.pk]*

[2]*Faculty of Engineering, Dalhousie University, Halifax, Canada.*

*Abstract*

**Advances in wireless communication, system on chip and low power sensor nodes allowed realization of Wireless Body Area Network (WBAN). WBAN comprised of tiny sensors, which collect information of patient's vital signs and provide a real time feedback. In addition, WBAN also supports many applications including Ubiquitous HealthCare (UHC), entertainment, gaming, military, etc. UHC is required by elderly people to facilitate them with instant monitoring anywhere they move around. In this paper, different standards used in WBAN were also discussed briefly. Path loss in WBAN and its impact on communication was presented with the help of simulations, which were performed for different models of In-Body communication and different factors (such as, attenuation, frequency, distance etc) influencing path loss in On-Body communications.**

## I. INTRODUCTION

With an increasing population around the world, specially the elderly people who are more fragile to health diseases, require a comprehensive HealthCare system. A system fulfilling needs of elderly people provides them with proper HealthCare facilities wherever, they move around. Wireless Body Area Network (WBAN) [1] is gaining attention worldwide for providing healthcare infrastructure. This system consists of several devices including tiny sensors which are placed in or around the body in close proximity to monitor a patient. As a result, elderly people are monitored everywhere and treated well intime in case of any emergency. The patients specially elderly people face problems in moving around and cannot frequently visit doctor(s), indeed require Ubiquitous HealthCare (UHC) [2].

For patients' monitoring, tiny sensors may be placed on or implanted in the body for constant monitoring. Different standards for WABN are defined which provide efficient means of data transfer and communication; Bluetooth, ZigBee, MICS and UWB [3]. A comprehensive and analytical survey is provided about these standards in this paper. Depending on the scenario of body communication (i.e., In-Body or On-Body), selection of antenna is very important, therefore, it has a direct effect on communication resulting in path loss.

When data collected by the sensors and devices is transferred through wireless medium to remote destination, then path loss is probable to occur. Path loss for In-Body and On-Body communications are different. It depends on frequency of operations as well as distance between transmitter and receiver. A simple path loss model for WBAN is proposed in [4].

We simulate In-Body path loss models proposed in [5] using MATLAB. In simulations, we considered four path loss models; deep tissue implant to implant, near surface implant to implant, deep implant to implant, and near surface implant to implant. Further, we perform simulations on different parameters effecting i.e., attenuation, phase distortion, RMS delay etc. communication in On-Body networks [6].

Rest of the paper is organized as follows: Section 2 discusses related work in WBAN specially in UHC. Section 3 describes the standards used in WBANs and the standard best suited for different architectures in UHC. In Section 4, path loss in WBAN and its effect in degradation of performance in communication are discussed in detail. Section 5 describes different scenarios of path loss in WBAN. Section 6 discusses the channel model and evaluation of different M-ary modulation schemes along with MATLAB simulation. Finally, Section 7 concludes the paper.

In next section, we briefly discuss the related work and motivation that paved way for this survey.

## II. RELATED WORK

Path loss models for medical implant and communication channels are presented by authors in [4]. They investigated statistical path loss model in Medical Implant Communication Service (MICS) channels. A work relating to path loss for On-body communications is provided by authors in [7].

In [8], sensor devices and server based architecture for UHC monitoring system is proposed. Introduction of wireless sensor devices and server part is given in this architecture. Communication between sensor and server is done via Base Station Transceiver (BST), which is connected to a server PC.

To provide services for the elderly people, components based system architecture of UHC monitoring is designed in [9]. A prototype system that monitors location and health status using Bluetooth as WBAN and smart phone with accelerometer as Intelligent Central Node (ICN) is used in this architecture. This architecture provides accessibility to family members or medical authorities to identify real time position and health status of patients via internet. ZigBee is used for small data rate applications because it consumes less power then Bluetooth.

WBAN architectures using wearable devices are proposed in [10]. One of them is wearable smart shirt, which is based on UHC and activity monitoring. This architecture comprises of smart shirt with multi-hop sensor network and server PC. Communication between smart shirt and server PC is done by BST. A device with two Process Control Block (PCB) mounted on each other is used in this architecture to reduce the size of integrated wearable sensor node along with Universal Serial Bus (USB) programming board as a separate module. It is needed only when nodes are connected to server PC.

One of the major issue in sensor networks is to utilize the energy efficiently. For utilizing the energy efficiently Mirak et. al. proposed distributed energy efficient algorithm specially for the problem of target coverage in [11]. For efficient energy utilization, sensors should be deployed in optimized manner. In [12], different sensor placements techniques are applied for analysing the efficiency of sensor network and also proposed an algorithm for solving the problem of complete coverage with minimum cost. Energy is one of the important factor affecting the performance of network. Even anchor placement scenerios and postion methodologies are also discussed in [13] for achieving the efficient energy utilization.

Recently, a lot of work is going on in the field of HealthCare and telemedicine. Wireless Body Sensors are being introduced, which provide efficient uses of resources. Sensors used in WBAN are lightweight, small in size, provide ultra-low power and are used for intelligent monitoring. These sensors continuously monitor human vital signs, physical activities and actions. There is increasing demand of UHC systems, which consume less power and provide longer battery lifetime. The system completely fulfills this demand. High data rate communication can be achieved by making body sensors compatible with underlying technologies. In this paper, a comprehensive and analytical survey is provided about the standards and devices for WBAN [1-4], [7-10],[14]. A detailed overview of UHC architectures in WBAN is provided. Also, we simulate In-Body path loss models proposed in [5] using MATLAB. In simulations, we considered four path loss models; deep tissue implant to implant, near surface implant to implant, deep implant to implant, and near surface implant to implant. Further, we performed simulations on different parameters effecting (i.e., attenuation, phase distortion, RMS delay etc) communication in On-Body networks [6]. In the next sectiopn, most frequently used WBAN standards are disscussed.

## III. MOST FREQUENTLY USED STANDARDS FOR WBAN COMMUNICATION

There is a number of standards which are adopted for communication in WBAN. Microscopic chips, which are typically used in wearable devices, depends on these standards. A detail discussion for standards; Bluetooth, ZigBee, MICS, and Ultra Wide Band (UWB) IEEE 802.15.6 [3].

### A. IEEE 802.15.1 (Bluetooth)

Bluetooth is a short range communication standard with data rate of 3 Mbps and range of about 10m. It is adopted in UHC due to high bandwidth and low latency. It also supports many mobile platforms. However, in UHC monitoring application, use of this standard is avoided because of high power consumption. It is suitable for latency and bandwidth sensitive scenarios [3].

### B. ZigBee

ZigBee standard is the most commonly used standard. It has the capability to handle complex communication in low power communication devices (such as, nodes) with collision avoidance schemes. It consumes less power (nearly 60 mW) and provides low data rate of 250 kbps. Hardware support with encryption is featured by many ZigBee controllers to provide effective protection for communication in WBAN [3].

### C. MICS

This band is specially designed for communication in WBAN. It is a short distance standard and is used to gather signals from different sensors on the body in a multi-hop structure. As compared to UWB, MICS has very low power radiation, thus, is most suitable for the sensors used in UHC monitoring system [3].

### D. IEEE 802.15.6 UWB

It provides very high bandwidth and data rate for communication. It is used for localization of transmitters. When very high bandwidth is required in any application, UWB is the best choice. For example, whenever, an emergency or critical situation occurs, UWB with Global Positioning System provides the best, short and traffic free route to the medical centre without any interference. User localization is usually important in hospitals or whenever, an emergency situation takes place. The advantage of UWB is that it is the only reliable method of localization. The drawback is receiver's complexity because of which it is not suitable for wearable applications in health monitoring [3].

After discussing the standards for WBAN, now we analytically study path loss in WBAN.

## IV. PATH LOSS IN WBAN

WBAN is greatly influenced by the amount of path loss that occurs due to different impairments. Devices for WBAN are generally placed inside or on the body surface, therefore, losses between these devices would affect the communication and can degrade the performance monitoring in UHC. In the following sections, we study in detail about WBAN communication and path loss that occurs in it and how it affects the performance of UHC.

Reduction in power density of an electromagnetic wave introduces path loss [4] [9]. Path loss is mainly caused by free space impairments of propagating signal like refraction, attenuation, absorption and reflection etc. It also depends on the distance between transmitter and receiver antennas, the height and location of antennas, propagation medium such as moist or dry air etc, and environment around the antennas like rural and urban etc [15]. Path loss for WBAN is different from traditional wireless communication because it depends on both distance and frequency. Frequency is catered because body tissues are greatly affected by the frequency on which sensor device is working.

Path loss model in $dB$ between transmitting and receiving antennas as a function of the distance $d$ is computed by [7] [16] as:

$$PL(d) = PL(do) + 10n\log_{10}(d/do) + \sigma_s \qquad (1)$$

where, $PL(do)$ is the path loss at a reference distance $d$, $n$ is the path loss exponent, and $\sigma_s$ is the standard deviation. Path loss in WBAN is of great importance. UHC in WBAN works well when the path loss between the transmitter and receiver is at its minimum. Path loss in WBAN occurs due to many factors such as reflection, diffraction and refraction etc, from the body parts which may distort the signal and can cause interference at receiver located at a distant location. So, data may face distortion due to path loss which causes difficulty for medical team located at far distance to correctly retrieve data. Path loss in UHC will decrease the efficiency of monitoring different vital signs in human body at patient's level as well as at medical team's level. The main focus of this section is to minimize the path loss that occurs at different stages in WBAN. This increases the efficiency of UHC monitoring in BAN which is our main goal. Path loss dependence on distance as well as frequency is given in Eq. (2) and Eq. (3):

$$PL = 20\log_{10}\left(\frac{4\pi d}{\lambda}\right) \qquad (2)$$

where, $L$ is the path loss in decibels, $\lambda$ denotes wavelength and $d$ specifies distance between transmitter and receiver [14].

As, we know that: $\lambda = \frac{c}{f}$; however, the above Eq. (2), can be rewritten as under:

$$PL = 20\log_{10}\left(\frac{4\pi df}{c}\right) \qquad (3)$$

*A.  WBAN*

In development of WBAN, one of the consideration is the characterization of the electromagnetic wave propagation from devices embedded inside the human body or close to it. A simple path loss model for WBAN is difficult to be driven in view of complex nature of human tissue structure and body shape. Since the antennas for WBAN application need to be placed on or inside the body, channel model has to take into consideration the effect of human body on radio propagation. To calculate the path loss in WBAN, three types of nodes are defined as under:

*a) Implant Node*
This type of node is embedded inside the body either below the skin or deeper.
*b) Body Surface node*
This type of node is placed on the surface of human skin or maximum 2cm away.
*c) External node*
This type of node is kept away from the body by a few centimeters upto a maximum of 5 meters.

For body surface communication, it is also important to consider distance between the transmitter and receiver around the body. If these are not placed on the same side in a straight line, then it allows creeping wave diffraction to be also taken into account. For external node communication, the distance between transmitter and receiver from the body vicinity is normally 3 meters away. However, in some cases, the maximum range of medical device can go upto 5 meters [17].

### B. Effect of WBAN Antennas

In case of antennas placed on the surface or inside the body, it is influenced by its surroundings. It is, therefore, essential to understand the changes in the antenna patterns and other characteristics must also be taken into account in the scenarios requiring propagation measurements. It is noticeable that the form factor of antenna is dependent on the requirements of applications. Different types of antennas are suitable for different applications e.g., for MICS, a circular antenna is used for pacemaker implant, while a helix antenna is most appropriate for a heart stent or urinary implant. Performance of antennas is greatly influenced by the form factor, which in turn effects the overall system performance. Antennas which take into account the characteristics of human body (such as, change in body tissues etc), are designed for measurements of channel model [18]. Antennas used in WBAN communication are categorized into following two types [19].

*a) Electrical antennas, such as dipole*

Electrical antennas are generally used for On-Body communications. They are avoided for In-Body communications, because, electromegnatic radiations of these antennas is harmful for tissues and muscles of body. On-Body communications, through these antennas do not make any direct contact with the body tissues and muscles; not resulting in heating of tissues.

*b) Magnetic antennas, such as loop*

Magnetic antennas are mostly used for In-Body and Implant communications. These antennas do not overheat the body tissues and is not dangerous to human body unlike electrical antennas. A loop of magnetic field is formed in magnetic antennas, which is within the defined range of the antenna, thus, these can communicate within this range not interfering with the body.

### C. Characteristics of Human Body

For wireless communication in WBAN, human body is not considered an ideal medium for the propagation of signal. Human body consists of materials which contain different dielectric constants, thickness and impedance which may not be ideal for communication. Depending on the frequency of operations, human body may encounter many impairments and losses such as absorption, attenuation and diffraction etc. Therefore, the characteristics of human body should be kept in mind before designing the path loss model for WBAN [20].

## V. SCENARIOS OF PATH LOSS IN WBAN

There are different scenarios of path loss which can take place in WBAN, since sensor nodes can be implanted inside the human body either planted on the surface of the body or atmost *2mm* away from the body surface. Since path loss is dependent on distance, as well as frequency, therefore, the variations of these parameters in these scenarios will affect the path loss model. To analytically examine these effects, we perform simulations using MATLAB. A detailed discussion of these simulations is given as under.

### A. In-Body Communication

In order to study propagation characteristics inside human body, simulations are carried out using a 3D visualization scheme [5]. The reason of using this scheme is that the study of physical parameters and their measurements are not feasible inside the human body. The antenna used in this study is a multi-thread magnetic loop antenna [21]. Like in [21], we consider four models for our simulations which include:
   a) Deep Implant to On-Body
   b) Near Surface Implant to On-body
   c) Deep Implant to Implant
   d) Near Surface Implant to Implant

Path loss calculations in oure simulations, are same as in [7] [16] (i.e., **Eq. (1)**) with a reference distance $d = 50mm$ and frequency of 402-405 MHz. The path loss exponent and standard deviation values for implant to body surface models are given in Table. 1.

Table. 1 Implant to Body Surface

| Models | Path Loss in $dB$ | $n$ | $\sigma_s$ ($dB$) |
|---|---|---|---|
| Deep Tissue Implant to Body Surface | 46.14 | 4.86 | 7.25 |
| Near Surface Implant to Body Surface | 47.81 | 4.532 | 6.23 |

Table. 2 Implant to Body Surface

| Models | Path Loss in $dB$ | $n$ | $\sigma_s$ ($dB$) |
|---|---|---|---|
| Deep Tissue Implant to Implant | 35.55 | 5.71 | 8.36 |
| Near Surface Implant to Implant | 41.25 | 5.12 | 8.95 |

From Table. 1 we conclude that the path loss at a reference distance for deep tissue implant to body surface is less than that of near surface implant to body surface because of high distance. Path loss exponent and standard deviation values for implant to implant models is given in Table 2.

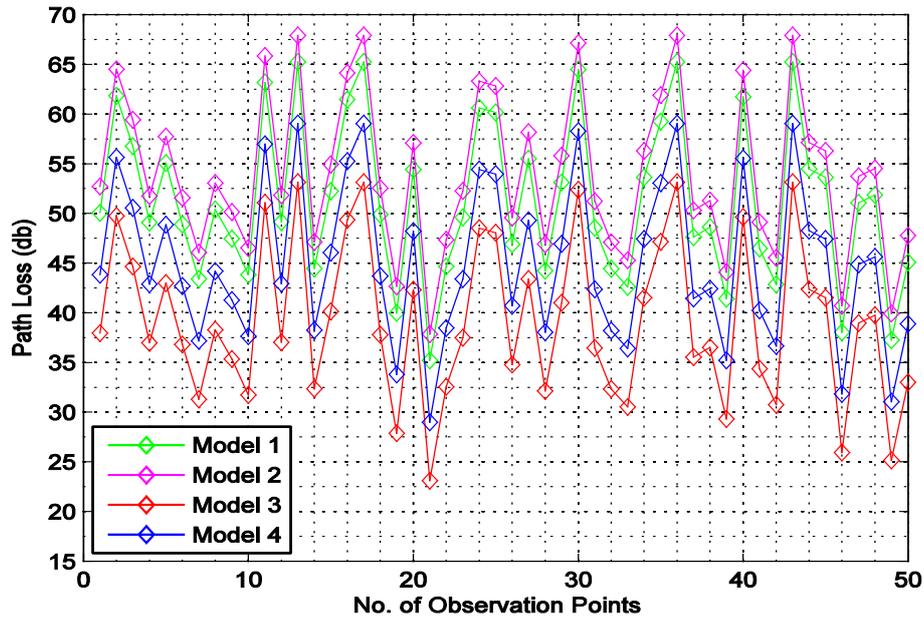

Fig. 2 Path Loss Models

Fig. 2 describes simulation results of path loss from deep tissue implant node to body surface node communication (Model 1). The simulation is carried out between no. of observation points between the deep tissue implant node and reference node placed at some distance, and path loss at each point of observation. The graph shows fluctuations in path loss at each observation point. The no. of observations are fixed at 50 for each model in the simulation. Curve of Model 1 in Fig. 2 is taken as reference for other three models. Since, no. of observation points are fixed at 50, therefore, the fluctuations are same for all models. However, increase or decrease in the path loss value is dependent upon the model which is being used. For near surface implant to body surface path loss model (Model

2), there is an increase of $4dB$ in path loss at each observation point from the reference model. For deep implant to implant and near surface implant to implant models, increase of $15dB$ and $9dB$ in path loss, respectively is noticed in Fig. 2. For Model 3, a decrease of $11dB$ in path loss occurs at each observation point from our reference model which is Model 1. The decrease is evident because distance between the nodes are smaller in this model (i.e., Model 3) from Model 1. With comparison to near surface implant to body surface path loss model, a decrease of $15dB$ in path loss is observed. This is because of further lessening of distance between the nodes, and decrease of $6dB$ in path loss from near surface implant to implant model. For near surface implant to implant path loss model (Model 4), a decrease of $5dB$ in path loss is obtained at each of the observation points from reference model (Model 1), $9dB$ for near surface implant to body surface path loss model (Model 2), while an increase of $6dB$ in path loss for deep tissue implant to implant path loss model (Model 3). If we further increase the no. of observation points, the fluctuations in path loss will be more sudden. This is due to different impairment factors such as refraction, diffraction, reflection etc. The summary of the models and their path loss with respect to the reference model is presented in Table 3.

Table. 3 Summary of In-Body Path Loss in WBAN

| Models | Results |
| --- | --- |
| Deep Tissue Implant to Body Surface (Model 1) | Reference Model |
| Near Surface Implant to Body Surface (Model 2) | Increase of 4dB, 15dB and 9dB in Path Loss from Model 1, 3 and 4, respectively |
| Deep Implant to Implant (Model 3) | Decrease of 11dB, 15dB and 6dB in Path Loss fron Model 1, 2 and 4, respectively |
| Near Surface Implant to Implant (Model 4) | Decrease of 5dB and 9dB in Path Loss from Model 1 and 2, respectively. Increase of 6dB in Path Loss from Model 3 |

    *B.       On-Body Communication*

For On-Body communications in WBAN, placement of sensors and actuators on the body surface is of great importance. Simple path loss model that takes into account the placement of sensors on the body, their communication with respect to body postures and movements are required. Channel response output of the On-Body communication as well as the frequency response can be easily found out. UHC monitoring in WBANs depend on both In-Body and On-Body communications of sensor nodes [6].

Amplitude attenuation of the signal with respect to frequency for On-Body communication is depicted in Fig. 3. As, frequency increases, attenuation of amplitude also increases since the channel undergoes impairments. Thus, these impairments degrades the intensity of signal, as it travels from transmitter to the receiver node planted on the human body.

Phase distortion of the signal with respect operating frequency for On-Body communication is obvious from Fig. 4. Direct relationship exists here as well; with increase in frequency the phase distortion of the signal increases in a linear fashion and vice versa.

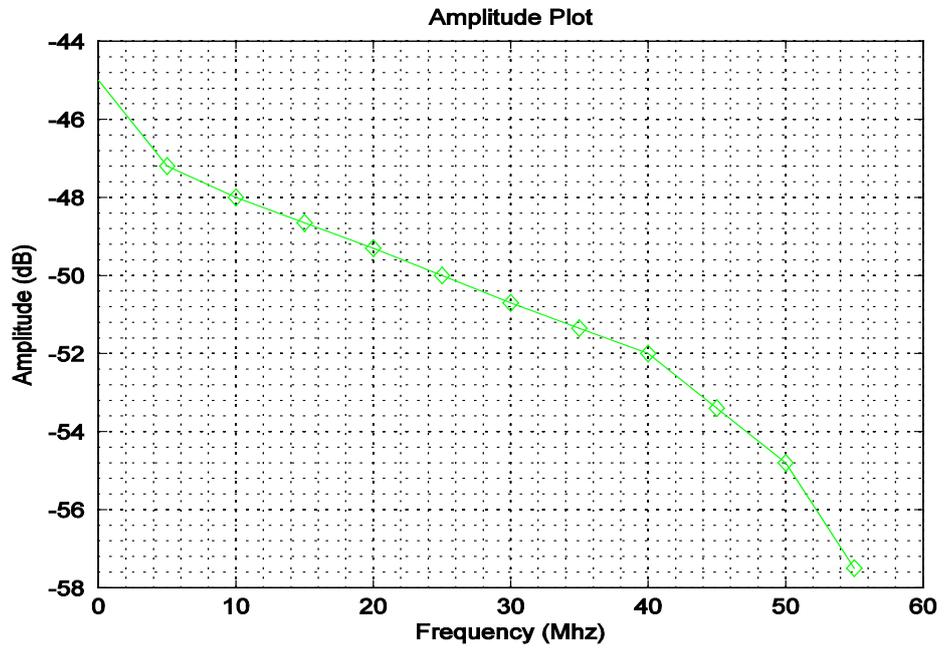

Fig. 3 Amplitude Attenuation in On-Body

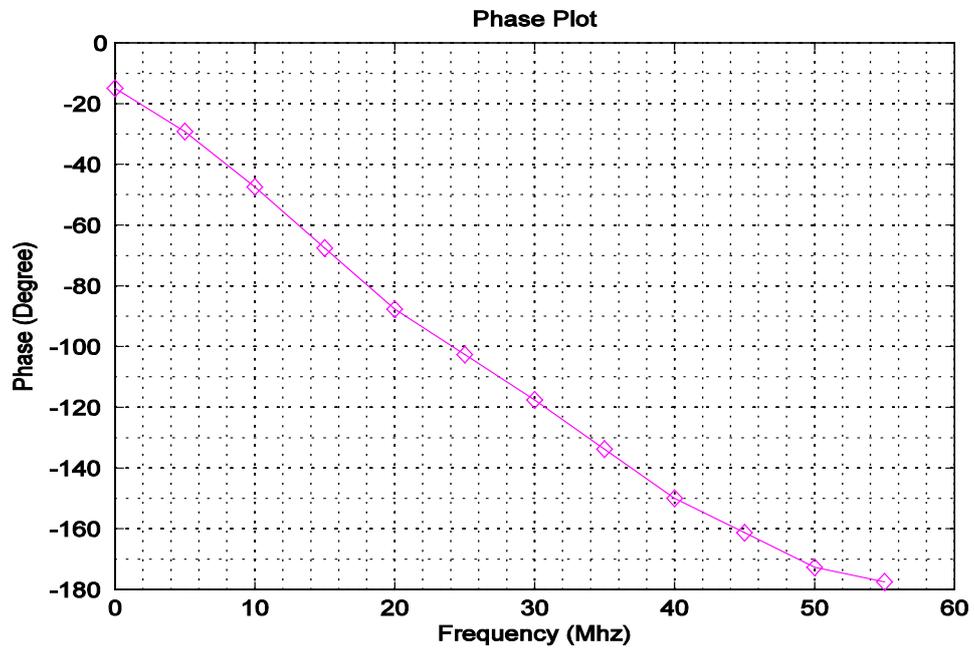

Fig. 4 Phase Distortion in On-Body

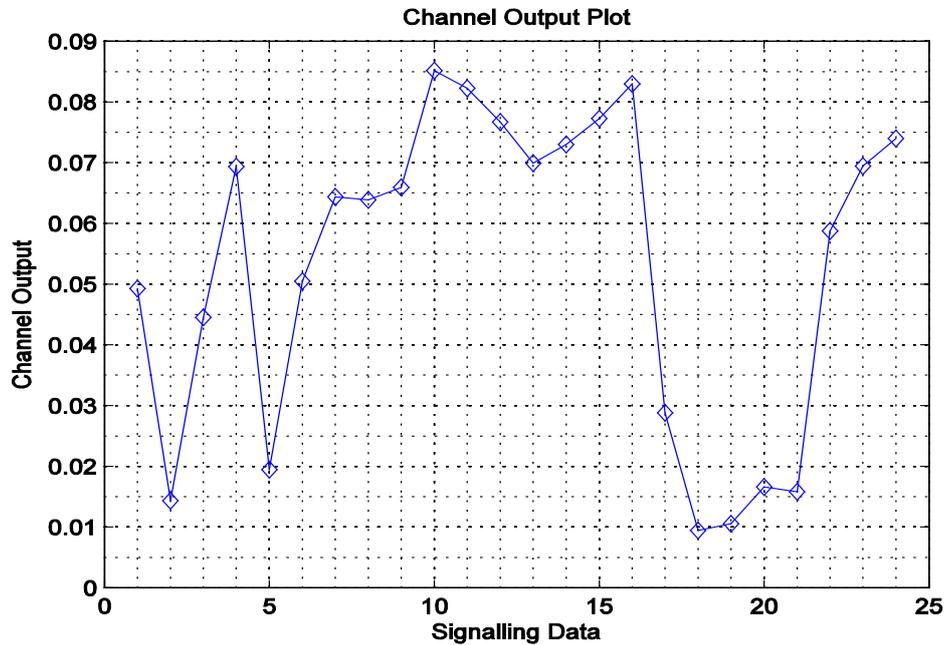

Fig. 5 Channel Output for On-Body Communication

Each component of the signal is distorted in phase and if relationship is not linear then there will be different phase distortion at different frequencies. From UHC point of view, both amplitude attenuation and phase distortion of the signal for On-Body communication should be eradicated to achieve better monitoring results.

Fig. 5 describes the channel output of On-Body communication with respect to the signalling data. The signalling data is uni-polar Non Return to Ground (NRG) stream of ones and zeros, respectively. Depending on this data, the channel output fluctuates with it having a higher output when the signalling data stream of more ones than zeros exists and lower output when the signalling data stream consists of more zeros than ones.

As, discussed earlier that path loss depends on the distance between the transmitting and receiving antenna/node as well as frequency of operation. Simulation results in Fig. 6 are carried out for two different frequencies; $900Mhz$ and $2.4Ghz$, as, it is obvious from Eq. (3) of path loss model. Since, direct relationship exists between path loss and distance between transmitter and receiver, therefore, by increasing distance, path loss increases linearly. Also, path loss has a direct relationship with frequency, thus, at $2.4Ghz$ the path loss curve is higher then that at $900MHz$.

Minimum delay spread is another requirement of On-Body communication and is analytically compared in our simulations. Delay spread or Root Mean Square (RMS) delay spread is a part of power delay profile.

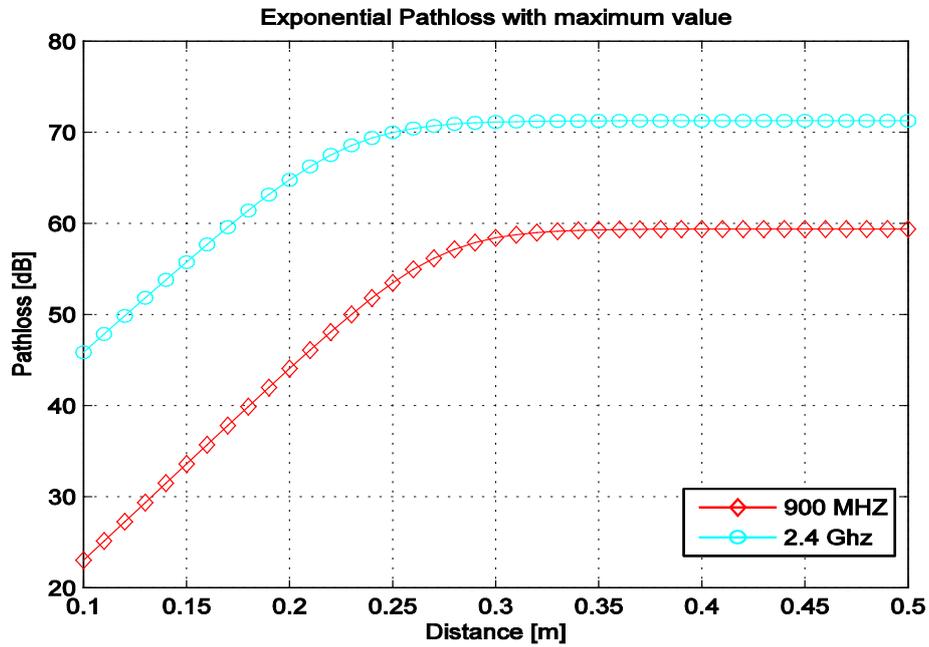
Fig. 6 Path Loss vs Distance for On-Body Communication

RMS delay spread as well as frequency dispersion of the signals operating at two different frequencies are determined by power delay profile. RMS delay spread is the standard deviation value of delay of reflections, weighted proportional to energy of signal [15].

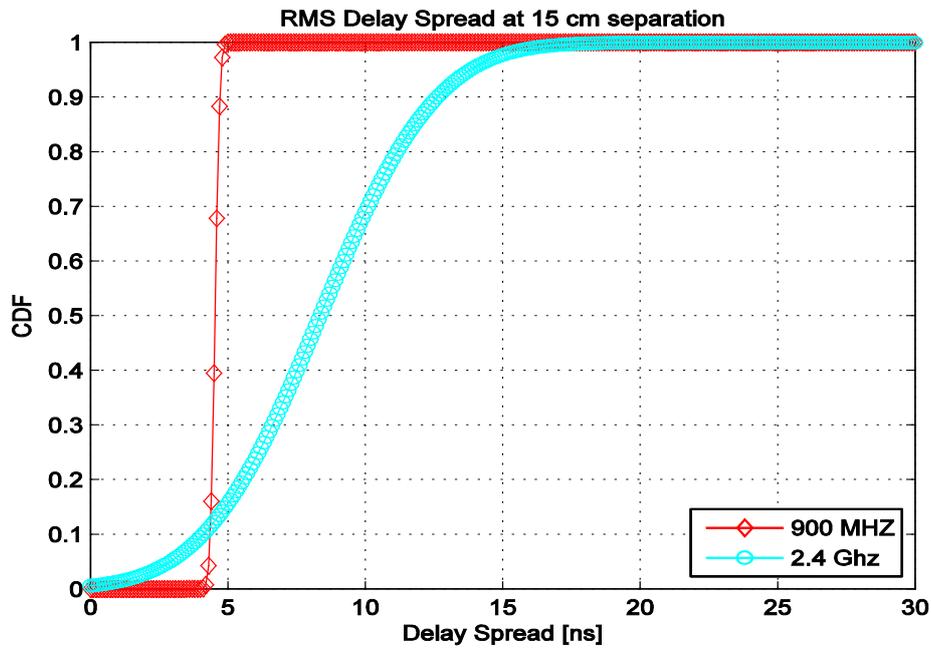
Fig. 7 RMS Delay at 15cm Separation

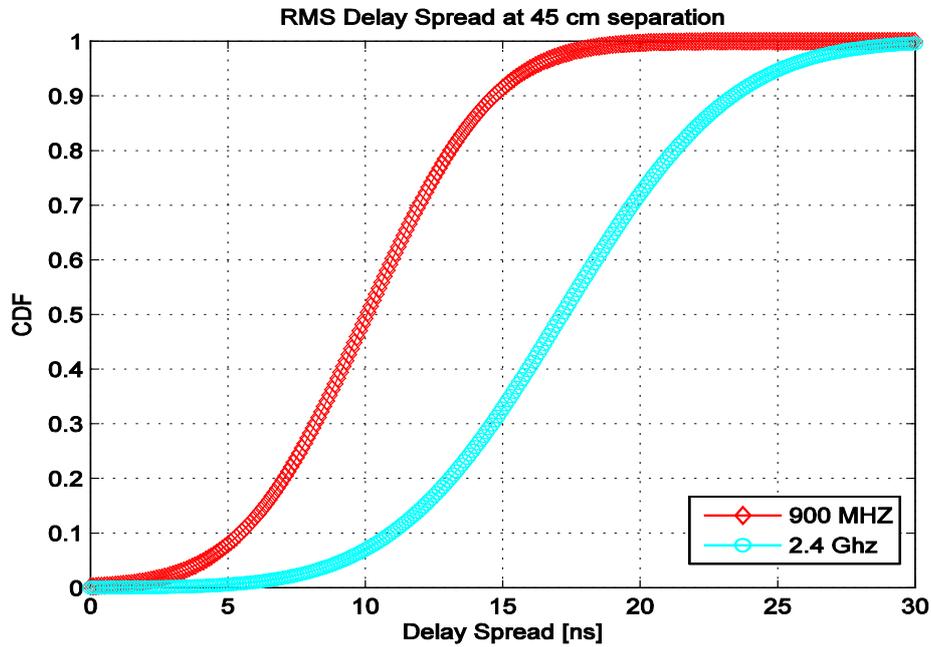

Fig. 8 RMS Delay at 45cm Separation

Fig. 7 and 8 show results of RMS delay spread measured at node separation of $15cm$ and $45cm$, respectively. Cumulative Distribution Function (CDF) is used to measure RMS delay spread and determines the probability of delay spread which occurs at a specified value. Thus, CDF value for probability of delay spread lies between 0 to 1. Graph shows that at $2.4Ghz$, RMS delay spread is having a slight linear curve, as compared to $900Mhz$ at which the delay spread is more straight. Fig. 8 portrays that if, we increase the distance then a straight relationship occurring at $900Mhz$ is changed into a curve similar to that is noticed at $2.4Ghz$. However, the delay spread at $900Mhz$ which is higher than that of $2.4Ghz$, because of correlation between the two frequencies.

The impairment factors and their results with respect to distance and frequency are summarized in Table 4.

Table. 4 Summary of On-Body Path Loss in WBAN

| Parameters | Results |
| --- | --- |
| Amplitude Attenuation | Increases with increase in frequency |
| Phase Distortion | Increases with increase in frequency |
| Path Loss | Increases with increase in distance and frequency |
| RMS Delay Spread (15cm Separation) | RMS delay spread increases initially for 2.4Ghz, however, in 900MHz RMS it is initially at 0 and then increases sharply |
| RMS Delay Spread (45cm Separation) | RMS delay spread increases initially for 900MHz and in 2.4GHz it takes some time |

## VI. WBAN Channel Model

In WBAN, radio propagations from devices that are close to or inside the human body are complex and distinctive comparing to the other environments, because the human body has a complex shape consisting of different tissues having their own permittivity and conductivity. Therefore, the channel models for WBAN are different from the ones in the other environments. Transmitter and Receiver are the integral part of WBAN channel. These models are already contributed to IEEE 802.15.6. The fading power profile of WBAN channel which includes fading and path losses are also discussed. In WBAN communications, propagation paths can experience fading due to different reasons, such as energy absorption, reflection, diffraction, shadowing by body, and body posture. The other possible reason for fading is multipath due to the environment around the body. Fading can be small scale or large scale. Small scale fading refers to the rapid changes of the amplitude and phase of the received signal within a small local area due to small changes in location of the on-body device or body positions, in a given short period of time. Whereas, large scale fading refers to the fading due to motion over large areas; this is referring to the distance between antenna positions on the body and external node (home, office, or hospital) [22].

### A. Evaluation of M-ary Modulations through a WBAN Channel

The error rate link performance has been evaluated and compared for efficient Rake receiver structures in low data rate WBAN channel. Rayleigh and Weibull distributions have been used for generating fading power profiles. This model is based on extensive measurements inside or on the surface of human body. The power profiles for WBAN channel have been generated. Bit Error Rate (BER) has been obtained using different M-ary modulation schemes; Quadrature Phase Shift Keying (QPSK), Binary Phase Shift Keying (BPSK), Frequency Shift Keying (FSK) and M-ary Multiple Shift Keying (MSK). With the aid of these graphs, bit error rate performance evaluation of rake receivers has been done. Through simulative investigations of BER, it has been found from these two modulations; MSK is suitable for WBAN channel. It has been observed that the performance of selective rake receiver (for optimum number of fingers) is better than partial rake receiver for all modulation schemes.

Since WBAN HealthCare applications require high data rate communication, so the channel model for WBAN should be less affected by the channel impairments. Different modulation schemes can be used in WBAN for obtaining high data rate signal at the receiver end. In our evaluation, we have taken into account rayleigh channel as well as Additive White Gaussian Noise (AWGN) channel for BER calculation which is done in MATLAB.

### B. Rake Receiver

In wireless applications, power consumption has been, and will be, one of the important characteristics when designing any wireless device. This is the case especially in sensor networks where a single sensor may be functioning, hopefully for very long time, without external power source. Generally, the architecture complexity reduces the battery life, however, performance increases with complexity. The best performance is achieved with the most complex devices which, however, consume a lot of power. Rake receivers can offer a good tradeoff between complexity and performance. In the near future, due to the aging of population, personal medical applications are most likely increasing in number and gaining more attention in industry. A rake receiver collects different multipath propagated signal components and coherently combines these in order to form a complete replica of a transmitted signal. There exist different complexity rake receivers. An all-rake (a-rake) sums up all the multipath components and in theory, collects all the signal energy. A-rake is the most complex and power consuming type of the rakes. Selective-rake (s-rake), on the other hand, collects the $n$ strongest multipath propagated signal components and sums them up for the detection. It is more realistic rake implementation than a-rake. However, both of the fore mentioned rake receivers require computationally expensive channel estimation and therefore are quite complex and power consuming rake types. S-rake needs calculation algorithm too for the decision of strongest signal components to be utilized.

A third rake type is a partial-rake (p-rake) where the $n$ first arriving signal components are processed at the receiver. This is the simplest rake receiver since synchronization and estimation of all the multi-path components are not required. Being less complex and less power consuming than either a-rake or s-rake, partial-rake can still be almost as good in performance [23]. In a Line-Of-Sight (LOS) channel, majority of signal energy is in the couple of tens of first arriving multipath components. Therefore the $n$ first taps only can be enough for a reliable decision. In our analysis and evaluation, we use p-rake receiver since it is less complex and less power consuming among selected recievers. The simulations for modulation schemes are done in MATLAB.

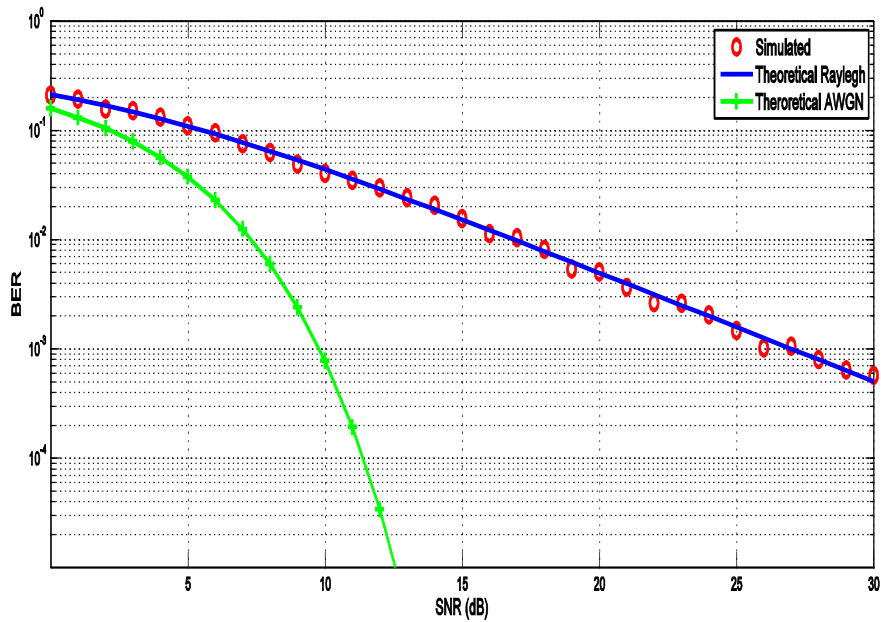

Fig. 9 BER Vs SNR of QPSK in Rayleigh Channel

Fig. 9 describes QPSK modulation for rayleigh channel used in WABN. It can be seen that maximum Signal-to-Noise Ratio (SNR) of $30 dB$ is achieved at BER close to 1. In case of MSK as shown in Fig. 10, SNR is less than that of QPSK, however, at a very low BER is better than QPSK.

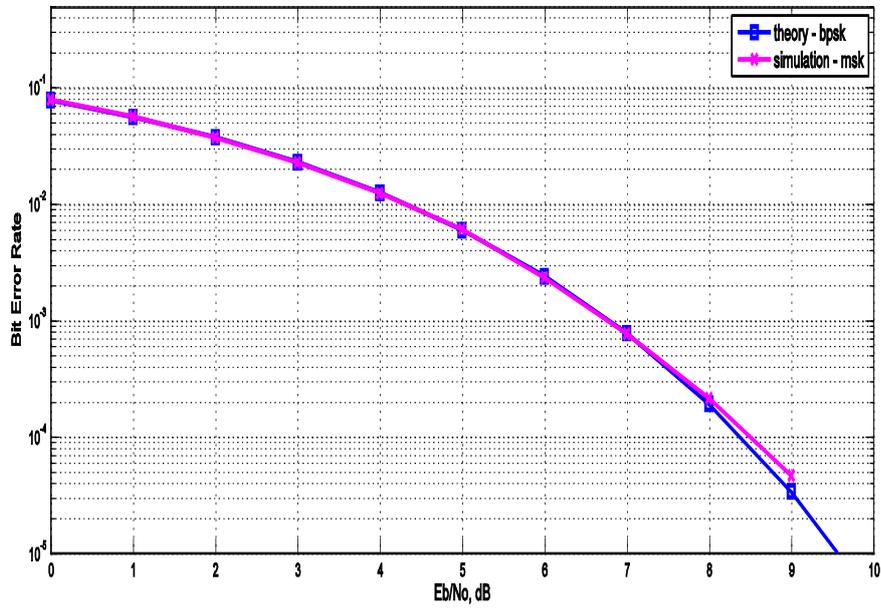

Fig. 10 BER Vs SNR of M-ary MSK in AWGN channel

Fig. 11 shows the simulation of FSK modulation along with its theoretical analysis which is exactly the same as simulated result. At $10dB$ SNR, BER is higher then that of MSK. But FSK performance is better then QPSK.

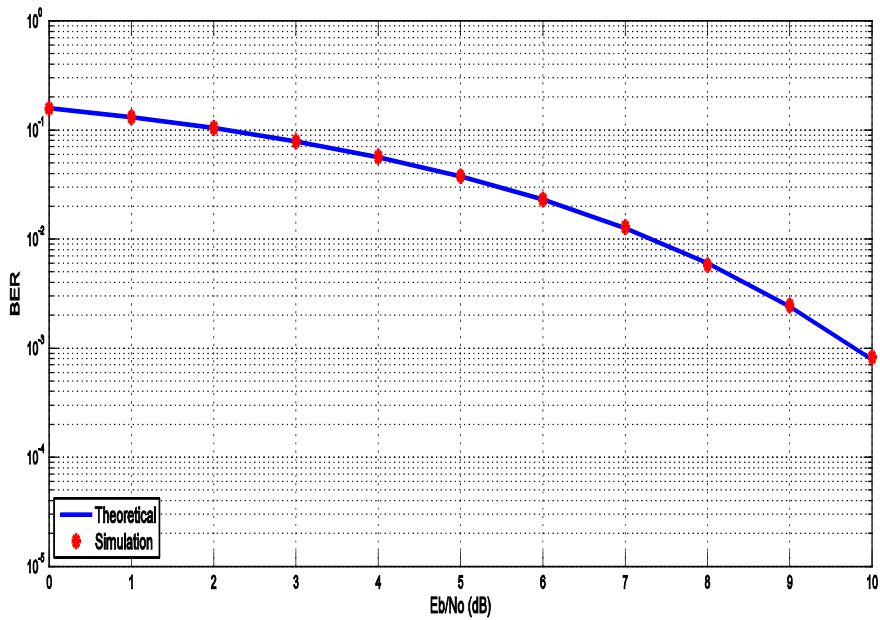

Fig. 11 BER Vs SNR of FSK in AWGN channel

In Fig. 12, BPSK simulation is shown along with theoretical result. It can be seen that BPSK performs best among all other modulation schemes used in WBAN channel. It is also compatible with p-rake receiver used for WBAN HealthCare application. It can be seen from the graph that at $10dB$ SNR, BER is very small which is desirable.

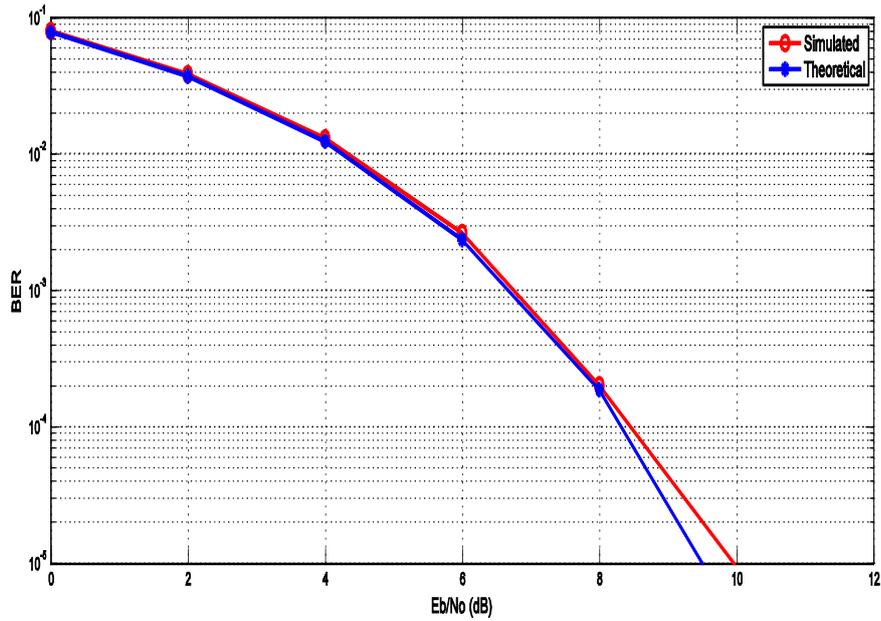

Fig. 12 BER Vs SNR of FSK in AWGN channel

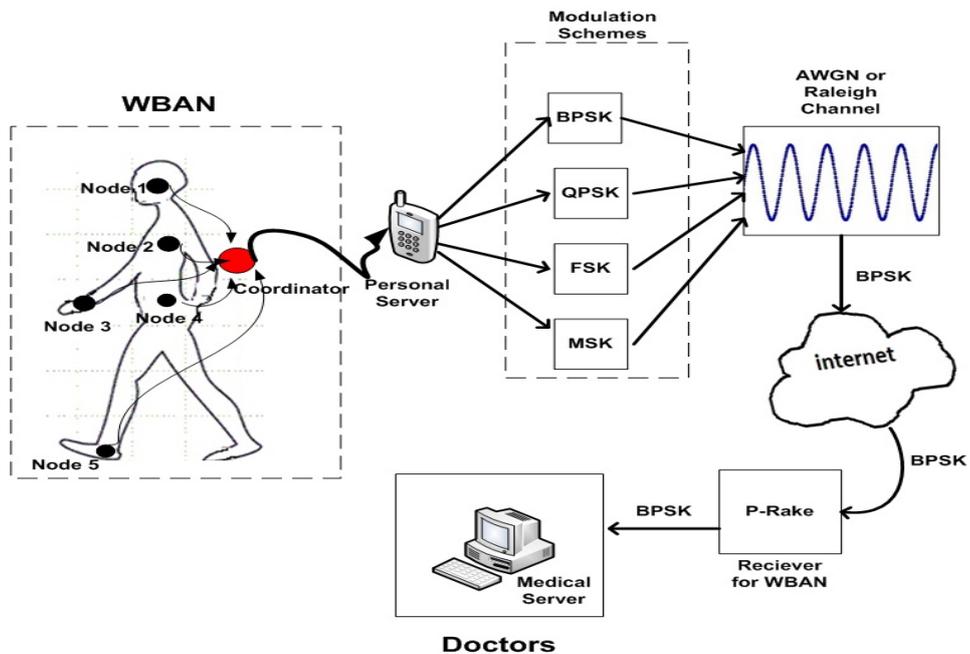

Fig. 13 Architecture of WBAN using Modulation Schemes

Fig. 13 describes the architecture of WBAN using the different M-ary modulation schemes discussed above. Since the performance of BPSK is best among others, therefore, the quality of signal using this schemes is good and also data rate of the communication is increased which is desired in WBAN HealthCare applications.

## VII. Conclusion

WBAN is an emerging domain in the field of wireless communication. It comprises of many tiny sensors placed on or inside the body. These sensors measure patient's vital information and transfer it to medical personnel for diagnosis. WBAN has many applications, most important of which is in UHC. With UHC, patients are not required to visit doctor frequently. They can get diagnosis and prescription of their disease while sitting at home. Nowadays, lot of work is going on to make low power sensors and devices that can be used in UHC. In this paper, different standards which are used in different types of WBAN applications are disscussed. Path loss in WBAN and its effects are also discussed in detail. Simulations for In-Body and On-Body communication are performed. The results for On-Body communications show that path loss increases between transmitter and receiver with increase in distance and frequency. Similarly, phase distortion and attenuation also increases with frequency. Moreover, path loss in different models of In-Body communication is also carried out. Finally, the channel model for WBAN is discussed with different modulation schemes suited for WBAN healthcare environment with their MATLAB simulations.